\documentclass[aip,amsmath,amssymb,
 reprint,%
]{revtex4-1}

\usepackage{graphicx,xcolor}% Include figure files
\usepackage{dcolumn}% Align table columns on decimal point
\usepackage{bm}% bold math
\usepackage[utf8]{inputenc}
\usepackage[T1]{fontenc}
\usepackage{mathptmx}
\usepackage{etoolbox}
\usepackage{amsmath}
\usepackage{amsthm}
\usepackage{braket} 
\usepackage{soul}
\usepackage{hyperref}

\makeatletter
\def\@email#1#2{%
 \endgroup
 \patchcmd{\titleblock@produce}
  {\frontmatter@RRAPformat}
  {\frontmatter@RRAPformat{\produce@RRAP{*#1\href{mailto:#2}{#2}}}\frontmatter@RRAPformat}
  {}{}
}%

\DeclareMathOperator{\Tr}{Tr}
\newcommand{\expect}[1]{\langle #1 \rangle}
\DeclareMathOperator*{\argmax}{arg\,max} %https://tex.stackexchange.com/questions/5223/command-for-argmin-or-argmax

\makeatother
\begin{document}

%\preprint{AIP/123-QED}

\title[Digital non-Gaussian characterization \& control]{Resource-efficient digital characterization and control of classical non-Gaussian noise}

\author{Wenzheng Dong}
\affiliation{\mbox{Department of Physics and Astronomy, Dartmouth College, Hanover, New Hampshire 03755, USA}}

\author{Gerardo A. Paz-Silva}%
%\email{Second.Author@institution.edu.}
\affiliation{Centre for Quantum Computation and Communication Technology (Australian Research Council), Centre for Quantum Dynamics, Griffith University, Brisbane, Queensland 4111, Australia} 

\author{Lorenza Viola}\thanks{
Corresponding author:
%Author to whom correspondence should be addressed: 
Lorenza.Viola@dartmouth.edu}
%\email{Lorenza.Viola@dartmouth.edu}
\affiliation{\mbox{Department of Physics and Astronomy, Dartmouth College, Hanover, New Hampshire 03755, USA}}

\date{\today}

\begin{abstract}
We show the usefulness of frame-based characterization and control [PRX Quantum {\bf 2}, 030315 (2021)] for non-Markovian open quantum systems subject to classical non-Gaussian dephasing. By focusing on the paradigmatic case of random telegraph noise and working in a digital window frame, we demonstrate how to achieve higher-order control-adapted spectral estimation for noise-optimized dynamical decoupling design. We find that, depending on the operating parameter regime, control that is optimized based on non-Gaussian noise spectroscopy can substantially outperform standard Walsh decoupling sequences as well as sequences that are optimized based solely on Gaussian noise spectroscopy. This approach is also intrinsically more resource-efficient than frequency-domain comb-based methods. 
\end{abstract}

\maketitle

Achieving accurate and predictive characterization and control (C$\&$C) of noise effects in qubit devices is essential for designing optimally-tailored, low-error quantum gates suitable for integration in quantum fault-tolerant architectures \cite{Cody_PRX,Khodjasteh_PRL_2009,Khodjasteh_PRA_2012,Caruso}. For many state-of-the-art scalable qubit platforms -- notably, solid-state qubits in superconducting circuits and semiconductor quantum dots -- noise is known to be {\em non-Markovian}, in the sense of exhibiting strong temporal correlations. While {\em non-Gaussian} noise statistics has also long been acknowledged to emerge in many scenarios of interest and lead to distinctive decoherence behavior \cite{Nakamura_echo,Paladino_RMP_2014}, renewed interest in non-Gaussian noise effects stems from both refined theoretical analyses \cite{Huang_2022,Dykman_2022} and recent experimental observations \cite{McCourt_DD_learn_noise,Rower_1f_noise}. Altogether, the non-Gaussian nature of the noise demands new methods for characterization and eventual error mitigation. 
 
Hamiltonian-level control techniques based on quantum noise spectroscopy (QNS) -- in pulsed \cite{Alvarez_Suter_RPL_2011,Yuge_PRL_2011,Szankowski_2017,Paz_mutilqubit_PRA_2017, Ferrie_2018,Paz_multiaxis_PRA_2019,Youssry_npj_2020,Barr_2022,Seif2023} or continuous-control modalities \cite{yan2013rotating,Frey_NatCom_2017,Norris_PRA_2018,Frey_PRApp_2020,Uwe2020} --  play a key role toward C\&C, allowing for noise statistical information (noise spectra or correlation functions) to be inferred from appropriately chosen control operations and measurement of system observables. Despite significant advances, standard QNS protocols suffer from several issues, however. On the one hand, they do not lend themselves to consistently incorporating the control constraints that are inevitably present in reality; as a result, the noise information provided by finite sampling must be supplemented by additional assumptions or approximations which need not be well justified. On the other hand, standard frequency-based QNS protocols are highly resource-inefficient, especially for non-Gaussian noise, whereby the estimation of {\em higher-order} noise correlations or spectra introduces significant extra challenges \cite{Norris_PRL_2016,Sung2019,Ramon_trispectrum_PRB_2019}.  

In this Letter, we develop a resource-efficient approach to {\em digital} C\&C for a qubit evolving under classical non-Gaussian noise. This is accomplished by leveraging a {\em control-adapted} (CA) description of the noisy qubit dynamics based on the notion of a {\em frame} \cite{Kovacevic2008Oct}, in which frame-based filter functions (FFs) and noise spectra are given a ``parsimonious'' representation directly tied to the finite control resources one can access~\cite{Teerawat_PRXQ_Frame,Wang_Walsh}. We focus qubit dephasing due to {\em random telegraph noise} (RTN)~\cite{Paladino_RMP_2014,Paladino_PRL_2002,FaoroViola,Abel_PRB_2008,Faoro2008,Paladino_PRL_2002,Sankar2008,Lisenfeld_NatCom_2015,LukeRTN,Cai_SciRep_2020,Chantasri_2022_RTN}, a realistic and ubiquitous non-Markovian classical noise model, and demonstrate how frame-based non-Gaussian QNS reconstructs spectral information relevant to qubit dynamics more efficiently than standard, comb-based non-Gaussian QNS can do. Moreover, we consider the task of designing noise-tailored dynamical decoupling (DD), and identify parameter regimes where control optimized on the basis of non-Gaussian QNS significantly outperforms standard digital (Walsh) DD schemes \cite{Hayes_walsh_PRA_2011} as well as control optimized solely on the basis of Gaussian QNS. 

Consider a qubit subject to dephasing from a classical environment and driven by time-dependent control. In a frame co-rotating with the qubit frequency, the relevant Hamiltonian may be written as $H(t)=\beta(t) \sigma_z + H_{\text{ctrl}}(t)$, where $\{\sigma_0\equiv {\mathbb I}, \sigma_u, u=x,y,z\}$ denote the Pauli basis, and we assume that $\beta(t)$ is a zero-mean, stationary classical stochastic process. By further assuming that control is perfect, the ideal Hamiltonian $H_0(t)\equiv H_{\text{ctrl}}(t)$. If $U_0(t)\equiv\mathcal{T}_+ e^{-i\int^t_0 ds\, H_{\text{ctrl}}(s)}$ (in units where $\hbar=1$ and with ${\mathcal T}_+$ denoting time ordering), moving to the interaction picture with respect to $H_{\text{ctrl}}(t)$ yields 
\begin{equation}
\widetilde{H}(t)= \beta(t) \sum_{u=x,y,z} y_{u}(t) \sigma_u , 
\end{equation}
where the ``switching functions'' $y_{u}(t)\equiv \tfrac{1}{2}\Tr[U^{\dagger}_0(t)\sigma_z U_0(t)\sigma_u]$ capture the effects of the applied control. The noise manifests in the measured time-dependent expectation values of qubit observables, $\langle O(T)\rangle \equiv {\mathbb E}\{ \Tr[{U}(T)\rho_0  {U}^{\dagger}(T) {O}] \}$, where $\rho_0$ is the initial qubit state, $U(t)\equiv\mathcal{T}_+ e^{-i\int^t_0 ds\, H(s)}$, $O=O^\dagger$, and ${\mathbb E}\{\cdot\}$ denotes the ensemble average over all the noise realizations. For simplicity, in what follows we will also use the symbol $\langle \cdot\rangle$ to denote ensemble averages. Assuming that $O$ is invertible, we may thus write \cite{Teerawat_PRXQ_Frame}
\begin{equation}
\label{VO}
\expect{O(T)} = \Tr [V_O(T) \rho_0 \widetilde{O}(T)], \quad
\widetilde{O}(T)\equiv U^{\dagger}_0(T)O U_0(T),
\end{equation}
in terms of a time-dependent (Hermitian) operator $V_O(T)\equiv \expect{\widetilde{O}^{-1}(T) \widetilde{U}^\dagger(T) \widetilde{O}(T)\widetilde{U}(T)}$, with $\widetilde{U}(T)=U_0^\dagger (T)U(T)$, that accounts for all the unwanted noise effects up to time $T$ \cite{Suppl}.

The operator $V_O(T)$ can be computed perturbatively, for instance, by means of a Dyson expansion, so that 
$\langle O(T) \rangle  = \Tr \,[\sum_{k=0}^\infty \mathcal{D}^{(k)}_O(T)/k! \,\rho_0 \widetilde{O}(T)].$
Let $f^u_c (T) \equiv -\tfrac{1}{2}\Tr[\widetilde{O}^{\dagger}(T)\sigma_u\widetilde{O}(T)\sigma_c]$. Then the $k$th-order Dyson contribution has the following structure \cite{Suppl}:
\begin{widetext}
\begin{equation}
      \frac{\mathcal{D}^{(k)}_O(T)}{k!}=(-i)^k\sum\limits^{k}_{l=0} \, \sum\limits_{\pi\in \Pi_{l;k}} \, \sum_{\vec{u},  \vec{c}} \,
    \int^T_0 d_>\vec{t}_{[k]}   \bigg[  \prod_{j=1}^l\prod_{j'=l+1}^k f^{c_j}_{u_{j}} (T ) \, y_{c_j}(t_{\pi(j)})y_{u_{j'}}(t_{\pi(j')})\sigma_{u_j}\sigma_{u_{j'}}  \bigg] 
       \expect{\beta(t_{1})\ldots \beta(t_{k})}, 
\label{eq:dyson_k_complicated}    
\end{equation}
\end{widetext}
with $\int^T_{-T}d_> \vec{t}_{[k]} \equiv \int^T_{-T} dt_1 \int^{t_1}_{-T} dt_2 \ldots \int^{t_{k-1}}_{-T} dt_k $ and $\Pi_{l;k}$ a set of index permutations. The term in the square bracket represents the control FF, whereas the noise properties enter through the $k$-point correlation function ($k$th-order moment). The above expansion identifies a dynamical integral $\mathcal{I}^{(k)}_{\vec{u}}(T) \equiv  \int^T_0 d_>\vec{t}_{[k]} \big[\prod_{j=1}^k y_{u_j}(t_j)  \big] {\mathcal L}(\vec{t})$, with ${\mathcal L}(\vec{t})$ a function of the noise correlations \cite{RemarkMagnus}. ``Orchestrating'' the controls to reveal noise correlation functions from a collection of $\expect{O(T)}$ is the fundamental spirit of QNS. Once the latter are known, the FFs can be designed in such a way to minimize these ``convolutions'', thus realizing noise-optimized control synthesis. 

More specifically, the objective of non-Gaussian QNS is to obtain information about the leading higher-order correlators (say, $k\leq K$), corresponding to the noise {\em polyspectra} in the frequency domain \cite{Brillinger, Norris_PRL_2016}. Since the expressions capturing the noise influence on the qubit dynamics hinge upon a perturbative expansion, $K$ is determined by requiring that the expansion remains accurate, up to a maximum evolution time $T$ of interest. Within a Gaussian approximation, $K=2$, the QNS task reduces to estimating the two-point correlator, $\expect{\beta(t_1) \beta(t_2)} = \expect{\beta(|t_1-t_2|) \beta(0)}$, whose frequency-domain Fourier transform is the well-known power spectral density (PSD)~\cite{Suppl}. Higher-order spectra, $K>2$, encode genuinely non-Gaussian noise features, whose %can have significant consequences in our ability to accurately control the target system. Their 
estimation comes at the cost of substantially higher protocol complexity~\cite{Norris_PRL_2016, Sung2019, Ramon_trispectrum_PRB_2019}.

However, learning the {\em full} form of the noise correlators or polyspectra is not only unnecessary, but also not a well-defined problem. 
In fact, given any kind of control constraints (e.g., maximum amplitude, finite pulse number, restricted pulse timings or control profiles), one may show that only certain ``components" of the noise correlations are relevant to the dynamics, and knowledge of this reduced set suffices for prediction and optimization of arbitrary controlled dynamics subject to the stated constraints. This is captured by the {\em frame-based CA FF formalism} introduced in Ref.~\onlinecite{Teerawat_PRXQ_Frame} which, as we will show, leads to a considerable complexity reduction in C\&C.

We focus on an scenario where control is restricted to sequences of $L$ instantaneous, equidistant perfect pulses over a time interval $[0,T]$, each pulse of the form $e^{i\theta \vec{n}\cdot \vec{\sigma}},$ with $\theta$ and $\vec{n}$ denoting an arbitrary rotation angle and direction, respectively. This yields piecewise-constant switching functions $y_u(t)$, corresponding to the widely used setting of {\em digital} control~\cite{Hayes_walsh_PRA_2011,Cooper, Haoyu,Wang_Walsh}. 
In the frame formalism, such digital switching functions can  be expanded in a ``window'' frame (in fact, a basis \cite{WalshR}), \mbox{$\mathcal{F}\equiv \{W_n(t) = \theta(t- n \tau) \theta ((n+1) \tau -t) \},$} with $\tau \equiv T/L$ and $n=\{1,\cdots,L\}$, such that   $y_u(t)=\sum_{n=1}^L F_u(n) W_n(t)$, with the expansion coefficient given by $F_u(n) = \frac{1}{\tau}\int^T_0 ds \,y(s)W_n(s).$  In this way, the relevant dynamical integrals take the form 
\begin{equation}
\begin{aligned}
\mathcal{I}^{(k)}_{\vec{u}}(T)|_{\text{CA}}&=\sum\limits_{\vec{n}}\prod\limits_{j=1}^k F^{(1)}_{u_j}(n_j)\bar{S}^{(k)}(\vec{n}), \\
 F^{(1)}_{u_j}(n_j)&\equiv\frac{1}{\tau}\int^T_0dt\, y_{u_j}(t){W}_{n_j}(t),\quad  j\in\{1,...,k\}, \\
   \bar{S}^{(k)}(\vec{n})& \equiv \int^T_0 d_>\vec{t}_{[k]} \expect{\beta(t_1)\cdots \beta(t_k)} \prod\limits^k_{j=1}W_{n_j}(t_j), 
\end{aligned}
\end{equation}
where $F^{(1)}_u(n)$ is identified as the \textit{frame-based FF}, while $\bar{S}^{(k)}(\vec{n})$ are the \textit{CA spectra}, with $1\leq n_k \leq \cdots n_1 \leq L$. In a CA QNS protocol, it is the CA spectra, which are  ``window-grained'' versions of the correlators $\expect{\beta(t_1)\cdots \beta(t_k)}$, that need to be estimated. This is in contrast to a standard, non-CA scenario, in which $\expect{\beta(t_1)\cdots \beta(t_k)}$ are needed for every $t_i$ (unless certain assumptions are made, as we shall discuss later), leading to the aforementioned complexity. For specified 
%(in our case, digital) 
control constraints, the complexity of obtaining the necessary and sufficient information to control the system to a given degree of accuracy in the CA vs. standard picture is dramatically different: if the size of the frame is taken as a measure for quantifying the resources needed for QNS, we have $\#\mathcal{C}^{(k)}_{\text{CA QNS}}\sim\mathcal{O}(L^k)$ while, in principle, $\#\mathcal{C}^{(k)}_{\text{full}}\sim\mathcal{O}(\infty)$ for full knowledge of the noise correlations. 

We now demonstrate how the frame-based CA approach affords resource-efficient C\&C in the presence of non-Gaussian RTN dephasing. In this case, $\beta(t) \equiv g \xi (t)$, where $g$ quantifies the noise strength and, for a symmetric RTN process switching with equal probability $\gamma$ between $\xi(t)=\pm 1$, the number of switches in $(0,t)$ is Poisson-distributed with mean $\gamma t$ \cite{Klyatskin2005}. The process is zero-mean stationary provided that $\xi(0) =\pm 1$ with equal probability, in which case the Gaussian and the leading non-Gaussian moments are 
\begin{equation}
    \begin{aligned}        
    \expect{\beta(t_1)\beta(t_2)}& = g^2 e^{-2\gamma (t_1 -t_2)} , \label{psd}\\
    \expect{\beta(t_1)\beta(t_2)\beta(t_3)\beta(t_4)} &= g^4e^{-2\gamma (t_1 -t_2+t_3-t_4)},
    \end{aligned}
\end{equation}
where $t_1 \geq t_2 \geq t_3 \geq t_4$. 
For fixed $\gamma$ and evolution time $T$, contributions from higher-order ($k>2$) $\beta$-cumulants
%, constructed from the $\xi$-cumulants 
are negligible for $g/\gamma \ll 1$, and the process is approximately Gaussian. Non-Gaussian features become prominent in a ``strong coupling'' (or ``slow fluctuator'') regime where $g/\gamma\gg 1$ \cite{Paladino_PRL_2002,Suppl}.  While $g$ is a time-independent parameter in the standard RTN model, we will also allow for the possibility of a coupling modulation of the form $g\mapsto g(t)\equiv g\cos(\Omega t +\phi)$, where $\Omega >0$ is a fixed but unknown value and $\phi$ is a random phase uniformly distributed in $[0,2 \pi]$. Physically, $\phi$ captures the uncertainty in the value of the coupling at $t=0$ when the experiment is started. In this way, it is possible to generate more general noise processes that have non-zero frequency features, while retaining stationarity~\cite{Suppl}.  

\begin{figure}
\centering    
\includegraphics[width=0.42\textwidth]{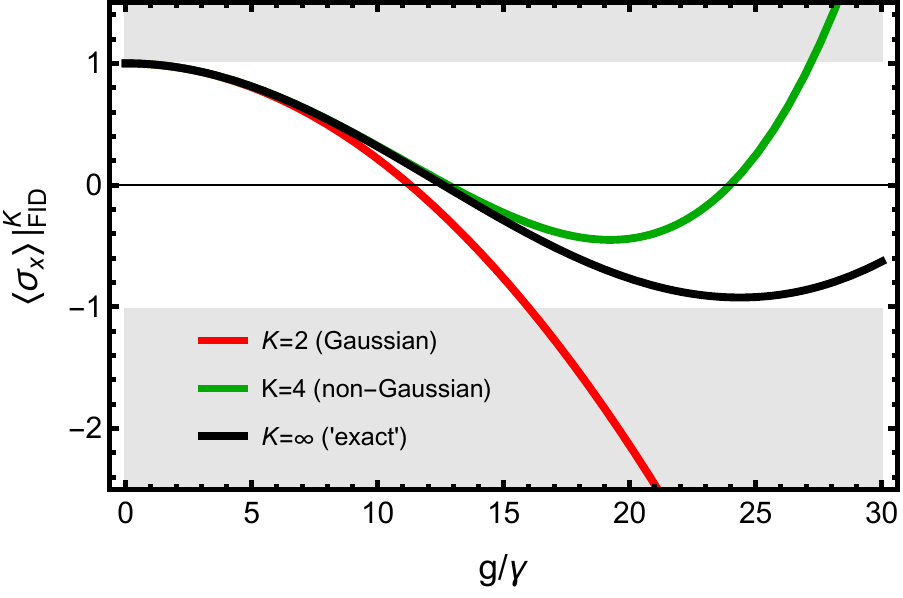}
\vspace*{-3mm}
\caption{{\bf Truncation error under free evolution.} Expectation values $\langle{\sigma_x}\rangle |^{K}_{\text{free}}$, with $\rho_0=\frac{1}{2}(\sigma_0+\sigma_y)$, under free evolution as a function of $g/\gamma$, for fixed time $T$ and different truncation orders, $K=2,4,\infty$. The shaded region is unphysical. Clearly, truncating at $K=4$ results in higher accuracy than $K=2$. In all our numerical simulations, we used RTN parameters $\gamma=0.02$ MHz, $T=3.2\,\mu$s. The intermediate coupling regime corresponds to $10<g/\gamma<40$.
}
\label{fig:obs_g_region_k_2_4_H}
\end{figure}

To evaluate the qubit dynamics under the above RTN noise, we truncate the expansion in Eq.\,\eqref{eq:dyson_k_complicated} at $K=4$. Accordingly, we work in an \textit{intermediate coupling regime} (in terms of $g/\gamma$), where it provides a good approximation of the dynamics and thus allows for an accurate CA spectra estimation, in the sense that (i) $\expect{ O(T)}  \approx \expect{ O(T)}|^K =\Tr \,[\sum_{k=1}^K \mathcal{D}^{(k)}_O(T)/k! \,\rho_0 \widetilde{O}(T)]$; and (ii) ${\widehat{\bar{S}}}{}^{(k\leq K)}(\vec{n})\approx \bar{S}^{(k\leq K)}(\vec{n})$. Fig.~\ref{fig:obs_g_region_k_2_4_H} provides a quantitative view of the error resulting from a truncation to order $K$ as a function of $g/\gamma$ for free evolution over a fixed time, indicating that a larger $K$ is required to access stronger coupling regimes. We note that free evolution is a worst-case scenario, as DD can extend the valid working regime of a truncation; for instance, we verified that $\langle \sigma_x\rangle \vert^{K=2}\approx 1$ over the entire range of couplings in Fig.~\ref{fig:obs_g_region_k_2_4_H}, if DD is applied.

\begin{figure*}
    \centering
    \includegraphics[width=.95\textwidth]{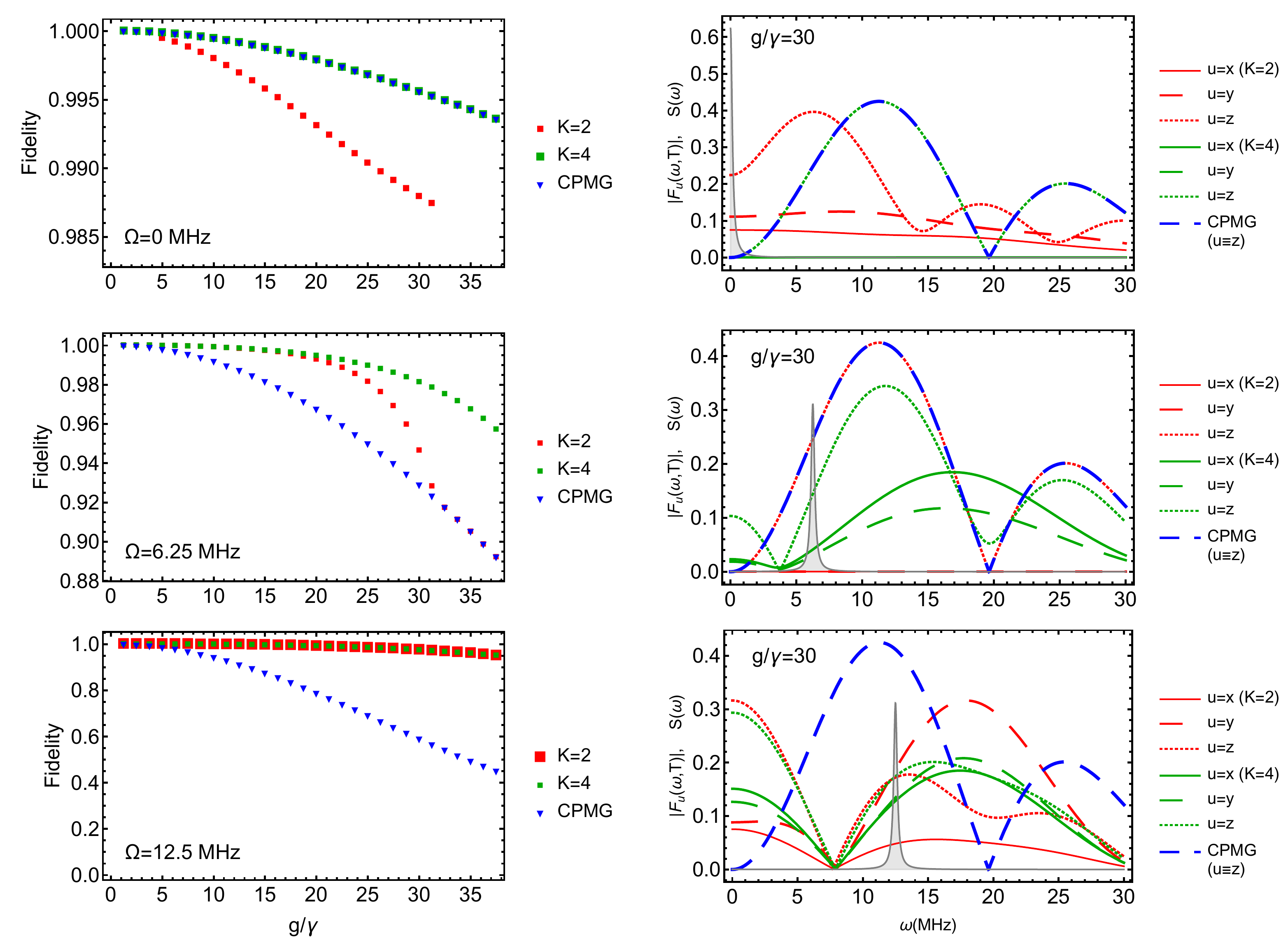}
    \vspace*{-3mm}
\caption{\textbf{Performance of noise-optimized control.} Process fidelity $F$ for three types of noise-optimized DD using a frequency-modulated stationary RTN model (ensemble size = 1,000, $\gamma T=0.064$) with different modulation frequency $\Omega$. Left column: fidelity against RTN coupling strength. Right column: RTN Lorentzian PSD  $S(\omega)$ and $|F_u(\omega,T)|$ ($u=\{x,y,z\}$) of the control, $P^*|^{K}_{\mathcal{F}}$, at $g/\gamma =30$. {\bf (Top)} For $\Omega=0$, the CPMG sequence has excellent filtering of zero-frequency spectra (note that peaks at $\pm \Omega$ merge in this case). The optimizer returns CPMG as the optimal solution, $P^*|^{K=4}_{\mathcal{F}} =P|_{\text{CPMG}}$. 
{\bf (Middle)} The PSD is not zero-centered; the non-Gaussian noise-tailored control significantly outperforms the other two DD schemes. As $g/\gamma \gtrsim 30$, $P^*|^{K=2}_{L=4} \rightarrow P|_{\text{CPMG}}$, reflecting the fact that large truncation errors offset any benefit of Gaussian QNS. Notably, the fidelity afforded by non-Gaussian CA QNS still exceeds the one of CPMG even at the highest coupling values \cite{Suppl}. 
{\bf (Bottom)} The strongly modulated PSD has a large overlap with the CPMG FF, making it performs poorly. In this case, however, the $K=2$ and $K=4$ QNS-optimized control have the same performance. 
}  
    \label{fig:QNS_optimized_gate_VS_CPMG}
\end{figure*}

Equipped with the frame expansion, we can express $\mathcal{D}^{(k\leq K)}_O(T)$ and $\expect{O(T)}|^K$ in the window frame analogously to $\mathcal{I}^{(k\leq K)}_{\vec{u}}(T)|_{\text{CA}}$. This yields algebraically complicated expressions, which we provide in the Supplement~\cite{Suppl}. Designing a CA QNS protocol $\mathcal{C}|_{\text{CA QNS}}$ then amounts to cycling over sufficiently varied pulse sequences (equivalently, different frame-based FFs ${F}^{(1)}_u(n)$), and over different initial qubit states $\rho_0$ and observables $O$, in such a way that the relevant CA spectra $\{\bar{S}^{(k)}(\vec{n}): 1\leq n_k\leq\cdots\ n_1 \leq L; ~1\leq k \leq K\}$ are learned. For $L=4$ as used in our simulation, the number of control settings needed for Gaussian vs. non-Gaussian CA estimation is $\#\mathcal{C}|^{K=2}_{\text{CA QNS}}=14$ vs. $\#\mathcal{C}|^{K=4}_{\text{CA QNS}}=49$. While we leave the details to the Supplement~\cite{Suppl}, our numerical simulation shows that not only can non-Gaussian CA QNS reconstruct the target CA spectra $\{\widehat{\bar{S}}{}^{(k\leq K)}(\vec{n})\}$ more precisely than Gaussian CA QNS does in the operating parameter regime, but it also leads to a significantly better prediction capability. 
 
Given knowledge of the CA spectra, our control objective is to craft a noise-optimized DD sequence (i.e., an identity gate) \cite{gate} over a fixed evolution period $T$. For every initial condition $\rho_0$, the time-evolved state of the qubit may be described as $ \rho(T)=\sum_{u,v=0}^3 \chi_{u v}(T) \sigma_u \rho_0 \sigma^{\dagger}_v,$ where the process matrix $\chi(T)$~\cite{OBrien_process_tomo_PRL_2004,Bialczak_NatPhy_2010} is a function of the control parameters $P$ (i.e., $\vec{n},~\theta$ of each pulse) and the inferred CA spectra. To obtain the optimized DD sequence, we use the knowledge of the numerically reconstructed CA spectra to maximize the process fidelity, $F (T)|^{K}_{\mathcal{F}} \equiv \Tr[\bar{\chi}(T) \widetilde{\chi}(T)] |^{K}_{\mathcal{F}}$, where $\bar{\chi}(T)$ ($\widetilde{\chi}(T)$) is the ideal (actual) $\chi$ matrix at time $T$. The subscript $\mathcal{F}$ stands for the $L=4$-window frame we are working in, whereas the truncation order $K$ indicates that only the knowledge up to the $K$th-order CA spectra is used in the numerical optimization. Concretely, the optimized digital control in such a frame, denoted  $P^*|^{K}_{\mathcal{F}}$, is determined by requiring that 
\begin{equation*}
        P^*|^{K}_{\mathcal{F}} \equiv \argmax_{P} F \Big(T; P|_{\mathcal{F}},\widehat{\bar{S}}{}^{(k\leq K)}(\vec{n}) \Big)\big|^{K}_{\mathcal{F}}.
\end{equation*} 
The maximum fidelity obtained in this way, $F(T; P^*|^{K}_{\mathcal{F}},\widehat{\bar{S}}{}^{(k\leq K)}(\vec{n}))|^{K}_{\mathcal{F}}$, does not account for the convolution between control and CA spectra beyond the truncation order ($\bar{S}^{(k > K)}(\vec{n})$). For comparison, the ``treu'' fidelity the optimized gate would deliver in experiment may be estimated by evaluating the performance of $P^*|^{K}_{\mathcal{F}}$ via exact numerical simulation of noise trajectories, resulting in $F(T;P^*|^{K}_{\mathcal{F}})\equiv F(T;P^*|^{K}_{\mathcal{F}},\bar{S}^{(k < \infty)}(\vec{n}))$.

In Fig.~\ref{fig:QNS_optimized_gate_VS_CPMG} we show the result of executing the above C\&C routine for Gaussian QNS-optimized control $F(T;P^*|^{K=2}_{L=4})$, non-Gaussian QNS-optimized control $F(T;P^*|^{K=4}_{L=4})$, and using standard Carr-Purcell-Meiboom-Gill (CPMG) $F(T;P|_{\text{CPMG}})$ as a benchmark, for three representative values of the RTN coupling modulation. Each left plot shows how the process fidelity between these three scenarios and the target identity operation varies as as function of $g/\gamma$, for noise profiles with dominant features at varying frequencies $\Omega$. For noise centered at $\Omega=0$ MHz, CPMG is essentially the optimal solution. As $\Omega$ grows, CPMG is rapidly outperformed by the optimal solution, which is to be expected given that ``plug-and-play'' routines such as DD sequences target low-frequency noise.  This can be explained by noticing that the shift in $\Omega$ changes the overlap between the relevant frequency-domain FF~\cite{Paz_PRL_2014}, given by $F_u(\omega,T) = \int_0^T ds\, e^{i \omega s} y_u(s)$, and the RTN PSD (two Lorentzian peaks centered at $\pm \Omega$) for $K=2$, and similarly in the higher-order multi-dimensional overlap integrals for $K>2$. The right side plots show the various functions entering the overlap integrals, providing an intuition about how different spectral features impact performance. Altogether, this demonstrates the superiority of noise-optimized controls under equivalent control constraints. 

The importance of characterizing non-Gaussian contributions to achieve high-fidelity operations becomes evident in the two scenarios with $\Omega >0$. For high modulation frequency, $\Omega=12.5$ MHz, the $K=2$ and $K=4$ characterizations yield the same optimal performance, implying that non-Gaussian noise features do not significantly contribute, regardless of the value $g/\gamma$. This is in stark contrast with the intermediate $\Omega=6.25$ MHz scenario. For small $g/\gamma,$ when the noise is effectively Gaussian~\cite{Bergli_NJP_2009}, the $K=2$ characterization is sufficient to achieve optimal performance. However, for larger $g/\gamma,$ when non-Gaussian contributions are expected to be more prominent, the $K=4$ characterization yields an optimal solution that considerably outperforms both the $K=2$-optimal solution and CPMG. Interestingly, this suggests that the ``Gaussianification'' of noise for small values of $g/\gamma$ is control-dependent, which complements existing results~\cite{Sankar2008}. Beyond the representative setting discussed here, the above demonstrates that characterizing high-order correlations can have a significant impact in optimizing gate implementations. 

\begin{figure}
\centering    
\includegraphics[width=0.43\textwidth]{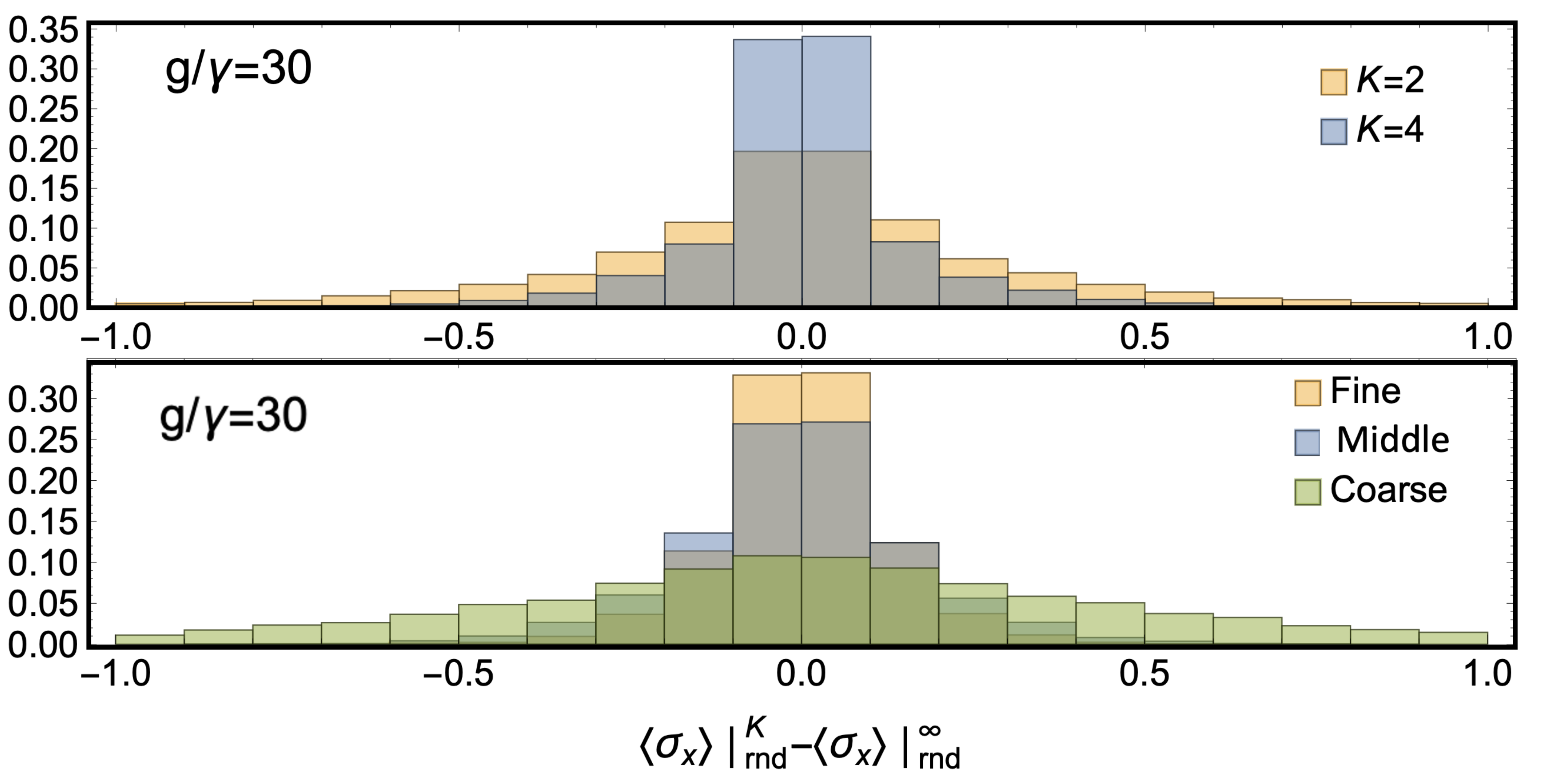}
\vspace*{-3mm}
\caption{{\bf Predictive capabilities under random control.} Distribution of prediction error,
%for an observable expectation value, 
$\expect{\sigma_x}|^K_{\text{rnd}}-\expect{\sigma_x}|^{\infty}_{\text{rnd}}$, under 10,000 random 4-window digital control realizations, each comprising four independent and random (in amplitude and phase) instantaneous pulses ($g/\gamma=30$, $\gamma T=0.064$,  $\rho_0=\frac{1}{2}({\mathbb I}+\sigma_y)$). {\bf (Top)} Error in CA QNS formalism with Gaussian ($K=2$) and non-Gaussian ($K=4$) truncation. The error stems only from truncation. {\bf (Bottom)} Error in the comb-method with non-Gaussian ($K=4$) truncation for three values of $\omega_0$: coarse ($\omega_0 =0.65$ MHz), middle  ($\omega_0 =0.42$ MHz), and fine resolution ($\omega_0 =0.13$ MHz). The error is due to both truncation and sampling density. A  performance comparable to the one of the CA method requires smaller $\omega_0$ to reduce the sampling error. 
}
\label{fig:histogram}
\end{figure}

One could argue that, given the narrow spectral structure in our model, the optimal solution requires minimal knowledge (e.g., knowing the position, not the height or shape, of the peak would suffice to achieve high fidelity); if so, optimal performance would not be a compelling argument for characterizing non-Gaussian correlations. A complementary  metric is shown in Fig.\,\ref{fig:histogram} (Top), illustrating how the $K=4$ characterization results in significantly better prediction -- a prerequisite for reliable optimization -- of the system behavior under random control sequences built from the admissible set.  The error in predicting a target $\sigma_x$ expectation value is below $0.1$ in $\sim 70\%$ of the cases for $K=4$, whereas the $K=2$ characterization  achieves this {\em only} in $\sim 40\%$ of the cases. We stress that the relatively high fraction of results with large error is not a limitation of the frame approach, but of the truncation order; better results can be achieved by a higher-$K$ characterization, 
at a higher cost ($\sim\mathcal{O}(L^K)$) in the necessary experiments.  

Having established and demonstrated the importance of characterizing higher-order noise correlations, we turn to assessing how the frame-based approach compares with established and experimentally demonstrated frequency-domain methods. Assuming stationarity, in frequency-domain QNS the aim is to reconstruct the Fourier transform of the noise cumulants, $C(\beta(t_1)\cdots \beta(t_k) )\equiv C(\beta(\tau_1)\cdots \beta(0))$, with $\tau_j\equiv t_j-t_k, ~j\in\{1,\cdots,k-1\}$, that is, to estimate the leading-order polyspectra $S_{k-1}(\vec{\omega})$. The PSD, bispectrum, and trispectrum correspond to $S_{1}({\omega})$ (Gaussian), and $S_{2}(\vec{\omega})$, $S_{3}(\vec{\omega})$ (non-Gaussian), respectively. Existing protocols capable of non-Gaussian QNS rely on a multi-dimensional {\it frequency-comb} approach~\cite{Norris_PRL_2016,Sung2019,Ramon_trispectrum_PRB_2019}, in which the repetition of a base sequence composed of $\pi$ pulses, say, $M\gg 1$ times, enforces the emergence of a comb structure in the FFs. In this case, the relevant dynamical integrals take the form~\cite{Suppl}
\begin{equation*}
\begin{aligned}
\mathcal{I}^{(k)}_{\vec{u}}(T)&|_{\text{freq}}=\int^{\infty}_{-\infty} d\vec{\omega}\big[\prod_{j=1}^{k-1}F(\omega_j,T) F(-\Sigma \vec{\omega},T)\big] S_{k-1}(\vec{\omega}) \\
\approx \frac{\omega_0^{k-1}}{M} & \!\!\sum_{ 
\substack{\Vec{m}, 
m_j\in \mathbf{Z}\\
\vec{m}\omega_0 \in \mathbb{D}_{k-1}}} \!\! \Big[\prod_{j=1}^{k-1}F(m_j\omega_0,T) F(-\omega_0\Sigma \vec{m},T)\Big]S_{k-1}(\vec{m}\omega_0), 
\end{aligned}    
\end{equation*}
where $\Sigma \vec{\omega}=\sum_{j=1}^{k-1}\omega_j$, the frequency resolution $\omega_0 = 2\pi/\tau_0$ is fixed by the length $\tau_0$ of the base sequence, and $\mathbb{D}_{k-1}$ denotes the {\em principal domain} of $S_{k-1}(\vec{\omega})$~\cite{Norris_PRL_2016,Chandran_1994Jan}. 

By executing experiments with different base sequences, one can sample the polyspectra at varying resolutions by using larger cycle times $\tau_0$. Since $S_{k-1}(\vec{\omega})$ are functions of continuous variables, continuous estimates are inferred by {\em interpolating} a finite set of estimates, $\{\widehat{S}_{k-1}(\vec{m}\omega_0)\}$ with, say, $|m_j| \leq m_{\text{max}} \equiv \lfloor \omega_{\rm max}/\omega_{0} \rfloor$, and  $\omega_{\rm max}$ the high-frequency cut-off of the noise. There is a clear trade-off between the accuracy of the interpolation and the sampling rate which, in turn, generates a trade-off between an accurate estimate of the polyspectra and the experimental resources needed for a desired sampling rate. Notice that there is also an implied smoothness assumption in these %frequency-domain 
protocols, as polyspectra with multiple narrow peaks would require a very small $\omega_0.$

Given this, it is possible to assess what sampling rate and experimental resources are necessary to match the predictive capability of the CA QNS. For RTN noise, $S_2(\vec{\omega})\equiv 0$, and the cost of sampling the PSD and the trispectrum is given by \cite{Ramon_trispectrum_PRB_2019}
\begin{equation*}
\begin{aligned}
    \#\mathcal{C}^{{K}=4}_{\text{comb-QNS}} &=  m_{\text{max}}+ \frac{1}{6} (m_{\text{max}}+1)(m_{\text{max}}+2)(2m_{\text{max}}+3).
    \end{aligned}
    \label{eq:principal_domain}
\end{equation*}
Table~\ref{tab:comb_QNS} shows the cost of sampling the polyspectra at four different resolutions, and the performance of the optimized control solution resulting from the corresponding interpolated polyspectra, for the scenario in Fig.\,\ref{fig:QNS_optimized_gate_VS_CPMG} (Middle). While the control performance is comparable in all cases, this is not a general feature; rather, it is due to the toy model of choice, for which minimizing the relevant overlap integrals is enough to locate the narrow peak, with other features being unimportant. All cases, however, yield an inferior optimal solution as compared to the CA QNS. The difference is most striking if one compares the prediction capability in each case with the CA one. Fig~\ref{fig:histogram} (Bottom) shows that only the high-resolution reconstruction comes close to the CA performance, while demanding $\sim 10^4$ more resources, thus demonstrating the power of the model-reduction in the CA formalism.  

\begin{table}
\begin{ruledtabular}
\begin{tabular}{l|lll}
%    \centering
%    \begin{tabularx}{0.48\textwidth}{|  l |X | X | X ||X| }
  %&  \shortstack{Min  \\ $\Delta \tau$ ($\mu s$)}   
  & \hspace*{-5mm}Min  $\Delta \tau$ ($\mu s$)
  &  \hspace*{-3mm}$\#\mathcal{C}|^{K=4}_{\text{QNS}}$ & \hspace*{-5mm}$F(P^*|^{K=4}_{\mathcal{F}/\omega}, T)$ \\
  \hline
  CA QNS & 0.2   & 49 &  98$\%$    \\
  \hline
     Comb-fine & 0.11   & 38,071 &  96.9$\%$   \\
            Comb-middle & 0.33  & 1,511  &96.6$\%$    \\
                Comb-coarse & 0.5  &  516 &  96.5$\%$      \\
    Comb-min & 1.3  &  59 &  95.8$\%$      \\
\hline
\end{tabular}
\end{ruledtabular}   
    \caption{\textbf{Comparison between comb-based and CA C\&C.} For precise reconstruction of the PSD/trispectrum, comb-based QNS requires much more resources than CA QNS, and shorter inter-pulse delay $\Delta \tau = T_{\text{coherence}}/ M/m_{\text{max}}$. In our estimation, we set  $T_{\text{coherence}}=100 \mu s$, $M=20$,   
    $m_{\text{max}}|_{\text{fine}}=47$, $m_{\text{max}}|_{\text{middle}}=15$, $m_{\text{max}}|_{\text{coarse}}=10$, and $m_{\text{max}}|_{\text{min}}=4$ (note that its  $\#\mathcal{C}|^{K=4}_{\text{comb QNS}}=59$ closely approximates $\#\mathcal{C}|^{K=4}_{\text{CA QNS}}=49$). $P^*|^{K=4}_{\omega}$ corresponds to the optimized 4-window DD (for $T=3.2\mu s$ as in the CA method) using non-Gaussian spectra reconstructed from comb-based QNS. Due to the narrow-band nature of the PSD/trispectrum, the optimized fidelity is insensitive to different frequency resolutions.   }
    \label{tab:comb_QNS}
\end{table}  

In conclusion, we leveraged the frame-based FF for digital control in a practically relevant non-Gaussian noise scenario. We showed that characterizing high-order noise correlations may be crucial to achieve the best possible gate, clearly outperforming plug-and-play control protocols. Furthermore, we demonstrated how exploiting the model-reduction capabilities of the CA formalism substantially reduces the resources needed for C\&C, 
%-- in this case, in a paradigmatic non-Gaussian classical noise setting -- 
making it experimentally more feasible. The generality of the formalism allows for a number of extensions, notably, to quantum non-Gaussian noise and non-instantaneous control. This is something we aim to explore for realistic scenarios in upcoming work.

\medskip

W.D. is deeply grateful to Yuanlong Wang for inspiring guidance and prompt material sharing, and to Muhammad Qasim Khan for valuable discussions. This work was supported by the U.S. Army Research Office through U.S. MURI Grant No. W911NF1810218 and by the Australian Government via AUSMURI Grant No. AUSMURI000002.

\iffalse
%%% LV: No need to show this in arXiv; I had to re-enter this info in the submission form...
\medskip

\begin{flushleft}
{\bf AUTHOR DECLARATIONS}\\
{\bf Conflict of Interest\vspace*{-2mm}}
\end{flushleft}

The authors have no conflicts to disclose.

\begin{flushleft}
{\bf Author Contributions\vspace*{-2mm}}
\end{flushleft}
% see https://publishing.aip.org/resources/researchers/policies-and-ethics/authors/

{\bf Wenzheng Dong:} 
Formal analysis (lead); 
Investigation (equal);
Software (lead);
Visualization (lead); 
Writing – original draft (lead); 
Writing – review \& editing (equal).
%
{\bf Gerardo A. Paz-Silva:} 
Conceptualization (supporting); 
Formal analysis (supporting); 
Funding acquisition (equal); 
Investigation (equal); 
Methodology (supporting);
Software (supporting);  
Writing – review \& editing (equal).
%
{\bf Lorenza Viola:} 
Conceptualization (lead); 
Formal analysis (supporting); 
Funding acquisition (equal);
Investigation (equal); 
Methodology (lead); 
Project administration (lead); 
Supervision (lead); 
Visualization (supporting);
Writing – original draft (supporting); 
Writing – review \& editing (equal).

\medskip
\begin{flushleft}
{\bf DATA AVAILABILITY\vspace*{-2mm}}
\end{flushleft}

The numerical data that support the findings of this study are available from the supplementary material and from the corresponding author on reasonable request.
\fi

\medskip

\bibliography{NGframes}

\begin{thebibliography}{54}
\bibitem [{gat()}]{gate}%
  \BibitemOpen
  \href@noop {} {}\bibinfo {note} {The frame formalism is by no means
  restricted to identity gates~\cite{Teerawat_PRXQ_Frame}. However, under the
  restriction of instantaneous control, it suffices to analyze the identity
  gate, as any other gate can be achieved by appending an appropriate $e^{-i
  \theta \vec{n}\cdot \vec{\sigma}}$-pulse after time $T$.}\BibitemShut {Stop}%
\end{thebibliography}

\end{document}